\documentclass[conference]{IEEEtran}
\IEEEoverridecommandlockouts

\usepackage{algorithmic}
\usepackage{amsmath,amssymb,amsthm}
\usepackage{bm} 
\usepackage{graphicx}
\usepackage{microtype}
\usepackage{pdfpages}
\usepackage{textcomp}
\usepackage{url}
\usepackage{xcolor}

\newtheorem{thm}{Theorem}[section]

\newtheorem*{prop*}{Proposition}
\theoremstyle{remark}

\def\BibTeX{{\rm B\kern-.05em{\sc i\kern-.025em b}\kern-.08em
    T\kern-.1667em\lower.7ex\hbox{E}\kern-.125emX}}
\usepackage[style=numeric,giveninits=true,backend=biber]{biblatex}
\bibliography{lostanlen2020eusipco.bib}

\begin{document}

\title{One or Two Frequencies?\\The Scattering Transform Answers
\thanks{This work is supported by the NSF award 1633259 (BIRDVOX).}
}

\author{\IEEEauthorblockN{Vincent Lostanlen}
\IEEEauthorblockA{\textit{Music and Audio Research Lab} \\
\textit{New York University}\\
New York, NY, USA\\
\url{vincent.lostanlen@nyu.edu}}
\and
\IEEEauthorblockN{Alice Cohen-Hadria}
\IEEEauthorblockA{
\textit{IRIT} \\
\textit{University of Toulouse --- CNRS} \\
Toulouse, France\\
\url{alice.cohen-hadria@irit.fr}}
\and
\IEEEauthorblockN{Juan Pablo Bello}
\IEEEauthorblockA{\textit{Music and Audio Research Lab} \\
\textit{New York University}\\
New York, NY, USA\\
\url{jpbello@nyu.edu}}
}

\maketitle

\begin{abstract}
With the aim of constructing a biologically plausible model of machine listening, we study the representation of a multicomponent stationary signal by a wavelet scattering network.
First, we show that renormalizing second-order nodes by their first-order parents gives a simple numerical criterion to assess whether two neighboring components will interfere psychoacoustically.
Secondly, we run a manifold learning algorithm (Isomap) on scattering coefficients to visualize the similarity space underlying parametric additive synthesis.
Thirdly, we generalize the ``one or two components'' framework to three sine waves or more, and prove that the effective scattering depth of a Fourier series grows in logarithmic proportion to its bandwidth.
\end{abstract}


\section{Introduction}
In the mammalian auditory system, cochlear hair cells operate like band-pass filters whose equivalent rectangular bandwidth (ERB) grows in proportion to their center frequency.
Given two sine waves $t \mapsto \bm{y_1}(t) = a_1 \cos(f_1 t + \varphi_1)$ and $t \mapsto \bm{y_2}(t) = a_2 \cos(f_2 t + \varphi_2)$ of respective frequencies $f_1 > 0$ and $f_2 > 0$, we perceive their mixture as a musical chord insofar as $\bm{y_1}$ and $\bm{y_2}$ belong to disjoint critical bands.
However, if $a_2 \ll a_1$ or $f_2 \approx f_1$, then the tone $\bm{y_2}$ is said to be \emph{masked} by $\bm{y_1}$.
In lieu of two pure tones, we hear a ``beating tone'': i.e., a locally sinusoidal wave whose carrier frequency is $\frac{1}{2} (f_1 + f_2)$ and whose modulation frequency is $\frac{1}{2} \vert f_1 - f_2 \vert$.
In humans, the resolution of beating tones involves physiological processes beyond the cochlea, i.e., in the primary auditory cortex.

The scattering transform ($\mathbf{S}$) is a deep convolutional operator which alternates constant-$Q$ wavelet decompositions and the application of pointwise complex modulus, up to some time scale $T$.
Broadly speaking, its first two layers ($\mathbf{S_1}$ and $\mathbf{S_2}$) resemble the functioning of the cochlea and the primary auditory cortex, respectively.
In the context of audio classification, scattering transforms have been succesfully employed to represent speech \cite{anden2014deep}, environmental sounds \cite{lostanlen2018jasmp}, urban sounds \cite{salamon2015eusipco}, musical instruments \cite{lostanlen2018dlfm}, rhythms \cite{haider2019cmmr}, and playing techniques \cite{wang2019ismir}.
Therefore, the scattering transform simultaneously enjoys a diverse range of practical motivations, a firm rooting in wavelet theory, and a plausible correspondence with neurophysiology.

This article discusses the response of the scattering transform operator to a complex tone input $\bm{y}:t \mapsto \bm{y_1}(t) + \bm{y_2}(t)$, depending on the sinusoidal parameters of $\bm{y_1}$ and $\bm{y_2}$.
In this respect, we follow a well-established methodology in nonstationary signal processing, colloquially known as: ``One or two frequencies? The X Answers'', where X is the nonlinear operator of interest.
The key idea is to identify transitional regimes in the response of X with respect to variations in relative amplitude ($\frac{a_2}{a_1}$), relative frequency ($\frac{f_2}{f_1}$), and relative phase ($\varphi_2 - \varphi_1$).
Prior publications have done so for X being the empirical mode decomposition \cite{rilling2008tsp}, the synchrosqueezing transform \cite{wu2011aada}, and the singular spectrum analysis operator \cite{harmouche2015gretsi}.
We extend this line of research to the case where X is the scattering transform in dimension one.

\section{Wavelet-based recursive interferometry}
Let $\boldsymbol{\psi} \in \mathbf{L}^2(\mathbb{R}, \mathbb{C})$ a Hilbert-analytic filter with null average, unit center frequency, and an ERB equal to $1/Q$.
We define a constant-$Q$ wavelet filterbank as the family $\boldsymbol{\psi}_{\lambda} : t \mapsto \lambda \boldsymbol{\psi}(\lambda t)$.
Each wavelet $\boldsymbol{\psi}_{\lambda}$ has a center frequency of $\lambda$, an ERB of $\lambda/Q$, and an effective receptive field of $(2\pi Q/\lambda)$ in the time domain.
In practice, the frequency variable $\lambda$ gets discretized according to a geometric progression of common ratio $2^{\frac{1}{Q}}$.
Consequently, every continuous signal $\bm{y}$ that is bandlimited to $[f_{\min}, f_{\max}]$ activates a number of $Q \log_2({\frac{f_{\max}}{f_{\min}}})$ wavelets $\boldsymbol{\psi_{\lambda}}$ at most.

We define the scalogram of $\bm{y}$ as the squared complex modulus of its constant-$Q$ transform (CQT):
\begin{equation}
    \mathbf{U_1}\bm{y} : (t, \lambda_1) \longmapsto
\left\vert \int_{-\infty}^{+\infty} \bm{y}(t^{\prime}) \bm{\psi_{\lambda_1}}(t - t^{\prime})\;\mathrm{d}t^{\prime} \right \vert^2.
\label{eq:scalogram}
\end{equation}
Likewise, we define a second layer of nonlinear  transformation for $\bm{y}$ as the ``scalogram of its scalogram'':
\begin{equation}
\mathbf{U_2} \bm{y} : (t, \lambda_1, \lambda_2) \longmapsto
\Big\vert \big \vert \bm{y} \ast \bm{\psi_{\lambda_1}} \big \vert^2 \ast \bm{\psi_{\lambda_2}} \Big \vert^2 (t),
\end{equation}
where the asterisk denotes a convolution product.
This construct may be iterated for every integer $m$ by ``scattering'' the multivariate signal $\mathbf{U_m}\bm{y}$ into all wavelet subbands  $\lambda_m < \lambda_{m-1}$:
\begin{multline}
    \mathbf{U_{m+1}} \bm{y} : (t, \lambda_1 \ldots \lambda_{m+1}) \longmapsto \\
\Big \vert \mathbf{U_{m}} \bm{y} (t, \lambda_1 \ldots \lambda_{m})
\ast \bm{\psi}_{\lambda_m} \Big \vert^2 (t, \lambda_1 \ldots \lambda_m).
\end{multline}

Note that the original definition of the scattering transform adopts the complex modulus ($\vert z \vert = \sqrt{z \bar{z}}$) rather its square ($\vert z \vert^2 = z \bar{z}$) as its activation function.
This is to ensure that $\mathbf{U_m}$ is a non-expansive map in terms of Lipschitz regularity.
However, to simplify our calculation and spare an intermediate stage of linearization of the square root, we choose to employ a measure of power rather than amplitude.
This idea was initially proposed by \cite{balestriero2017arxiv} in the context of marine bioacoustics.

Every layer $m$ in this deep convolutional network composes an invariant linear system (namely, the CQT) and a pointwise operation (the squared complex modulus).
Thus, by recurrence over the depth variable $m$, every tensor $\mathbf{U_m} \bm{y}$ is equivariant to the action of delay operators.
In order to replace this equivariance property by an invariance property, we integrate each  $\mathbf{U_m}$ over some predefined time scale $T$, yielding the invariant scattering transform:
\begin{align}
    \mathbf{S_{m}} \bm{y} : (t, p) \longmapsto
    \int_{-\infty}^{+\infty} \mathbf{U_{m}}(t^{\prime}, p) \bm{\phi}_{T}(t-t^\prime)\;\mathrm{d}t^{\prime},
\end{align}
where the $m$-tuple $p = (\lambda_1 \ldots \lambda_m)$ is a scattering path and the signal $\bm{\phi}_T$ is a real-valued low-pass filter of time scale $T$.

\section{Auditory masking in a scattering network}

\begin{figure}
    \begin{center}
    \includegraphics[width=0.95\linewidth,keepaspectratio]{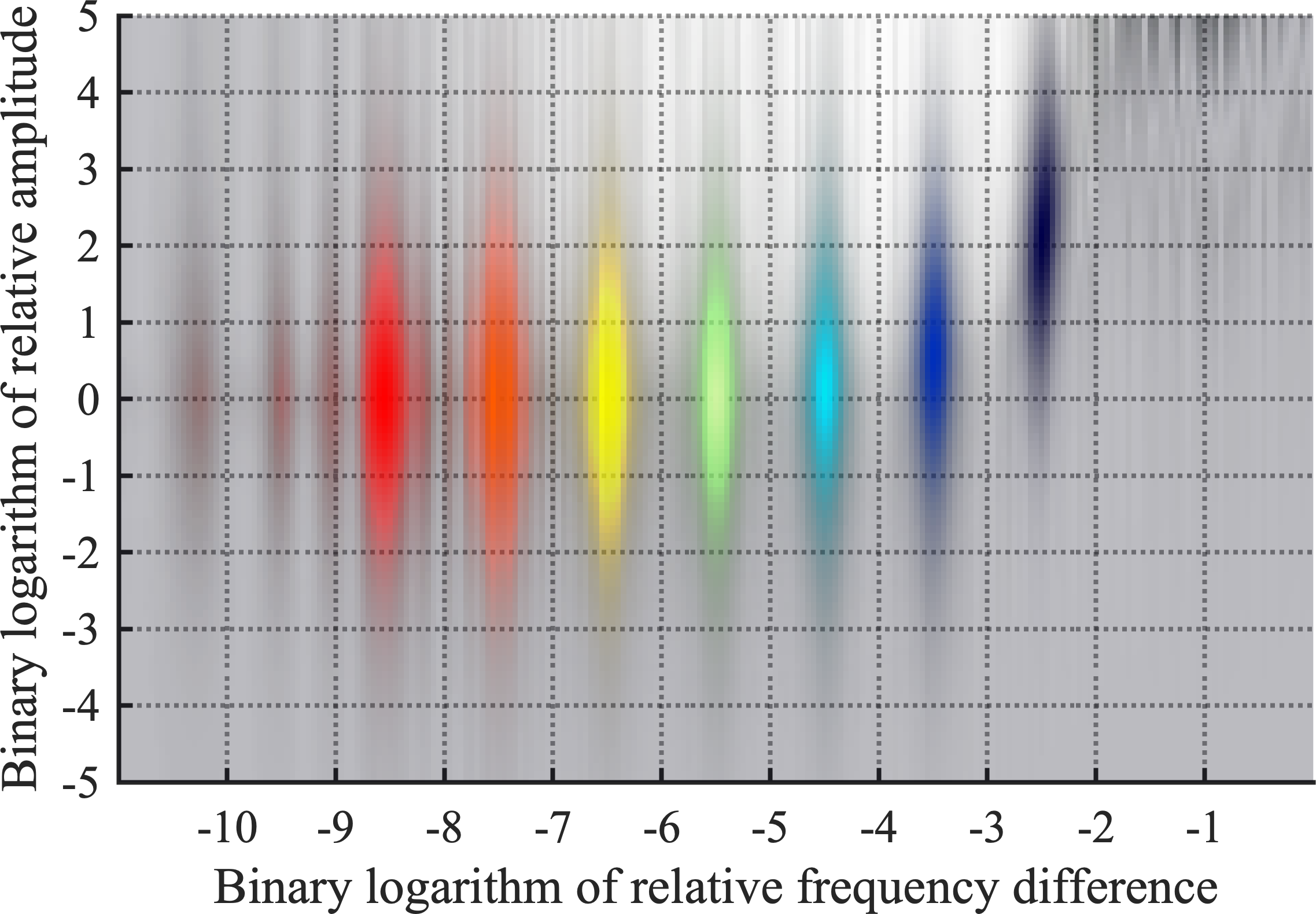}
    \end{center}
    \caption{
    Superimposed heatmaps of second-order masking coefficients $\mathbf{\widetilde{S}_2}\bm{y}$ after a scattering transform of two sine waves $\bm{y_1}$ and $\bm{y_2}$, measured around the frequency $f_1$, as a function of relative amplitude $\frac{a_2}{a_1}$ and relative frequency difference $\frac{\vert f_2 - f_1 \vert }{f_1}$.
    The color of each blot denotes the resolution $\lambda_2$ at the second layer.
    Wavelets have an asymmetric profile (Gammatone wavelets) and a quality factor $Q=4$.
    The second layer covers an interval of nine octaves below $f_1$.
    For the sake of clarity, we only display one interference pattern per octave.
    \label{fig:heatmap}
    }
\end{figure}

Given $n\in\{1, 2\}$, the convolution between every sine wave $\bm{y_n}$ and every wavelet  $\bm{\psi_{\lambda_1}}$ writes as a multiplication in the Fourier domain.
Because $\bm{\psi_{\lambda_1}}$ is Hilbert-analytic, only the analytic part $\bm{y_n^{\mathrm{a}}} = \bm{y_n} + \mathrm{i} \mathcal{H}\{\mathbf{y_n}\} = a_n \exp(\mathrm{i}(f_n t + \varphi_n))$ of the real signal $\bm{y_n}$ is preserved in the CQT:
\begin{equation}
    \left(\bm{y_n} \ast \bm{\psi_{\lambda_1}}\right)(t)
    = \dfrac{1}{2} \bm{\widehat{\psi}_{\lambda_1}}(f_n) \bm{y_n^\mathrm{a}}(t).
\end{equation}
By linearity of the CQT, we expand the interference between $\bm{y_1}$ and $\bm{y_2}$ by heterodyning:
\begin{align}
    & \Big\vert ( \bm{y_1} + \bm{y_2} ) \ast \bm{\psi_{\lambda_1}} \Big\vert^2 (t) =
    \frac{1}{2}  \Big\vert\bm{\widehat{\psi}}\big(\frac{f_1}{\lambda_1}\big) \Big\vert^2 a_1^2
    + \frac{1}{2}  \Big\vert\bm{\widehat{\psi}}\big(\frac{f_2}{\lambda_1}\big) \Big\vert^2 a_2^2
    \nonumber \\ &
    + \mathfrak{R}\Bigg(\!\bm{\widehat{\psi}}\Big(\frac{f_1}{\lambda_1}\Big) \bm{\widehat{\psi}^\ast}\Big(\frac{f_2}{\lambda_1}\Big)\!\Bigg) a_1 a_2 \cos\big((f_2 - f_1) t + (\varphi_2 - \varphi_1) \big).
    \label{eq:de-deux-composantes-a-une}
\end{align}
Because the wavelet $\bm{\psi}$ has a null average, the two constant terms in the equation above are absorbed by the first layer of the scattering network, and disappear at deeper layers.
However, the cross term, proportional to $a_1 a_2$, is a ``difference tone'' of fundamental frequency $\Delta\!f = \vert f_2 - f_1 \vert$.

The authors of a previous publication \cite{anden2012dafx} have remarked that this difference tone elicits a peak in second-order scattering coefficients for the path $p=(\lambda_1, \lambda_2)=(f_1, \vert f_2 - f_1 \vert)$.
In the following, we generalize their study to include the effect of the relative amplitude $\frac{a_2}{a_1}$, the wavelet shape $\bm{\psi}$, the quality factor $Q$, and the time scale of local stationarity $T$.

Equation \ref{eq:de-deux-composantes-a-une} illustrates how the scalogram operator $\mathbf{U_1}$ converts a complex tone (two frequencies $f_1$ and $f_2$) into a simple tone (one frequency $\vert f_2 - f_1 \vert$).
For this simple tone to carry a nonnegligible amplitude in $\mathbf{U_2}$, three conditions must be satisfied.
First, the rectangular term $a_1 a_2$ must be nonnegligible in comparison to the square terms $a_1^2$ and $a_2^2$.
Secondly, there must exist a wavelet $\boldsymbol{\psi}_{\lambda_1}$ whose spectrum encompasses both frequencies $f_1$ and $f_2$.
Said otherwise, $\lambda_1$ must satisfy the inequalities $\vert \frac{f_n}{\lambda_1} - 1 \vert \ll \frac{1}{Q}$, both for $f_n = f_1$ and for $f_n = f_2$.
Thirdly, the frequency difference $\vert f_2 - f_1 \vert$ must belong to the passband of some second-order wavelet $\bm{\psi_{\lambda_2}}$.
Yet, in practice, to guarantee the temporal localization of scattering coefficients and restrict the filterbank to a finite number of octaves, the scaling factor of every  $\bm{\psi_{\lambda_m}}$ is upper-bounded by the temporal constant $T$.
Therefore, the period $\frac{2\pi}{\vert f_2 - f_1\vert}$ of the difference tone should be under the pseudo-period of the wavelet with support $T$; i.e., a pseudo-period of $Q T$.
Hence the third condition: $\vert f_2 - f_1 \vert \ll \frac{2 \pi Q}{T}$.

One simple way of quantifying the amount of mutual interference between signals $\boldsymbol{y_1}$ and $\boldsymbol{y_2}$ is to renormalize second-order coefficients by their first-order ``parent'' coefficients:
\begin{equation}
    \mathbf{\widetilde{S}_2}\bm{y}(t,\lambda_1,\lambda_2) = \dfrac{\mathbf{S_2}\bm{y}(t, \lambda_1, \lambda_2)}{\mathbf{S_1} \bm{y}(t, \lambda_1)}
\end{equation}
This operation, initially proposed by \cite{anden2014deep}, is conceptually analogous to classical methods in adaptive gain control, notably per-channel energy normalization (PCEN) \cite{lostanlen2019spl}.

In accordance with the ``one or two frequencies'' methodology, Figure \ref{fig:heatmap} illustrates the value of this ratio of energies in the subband $\lambda_1 = f_1$, for different values of relative amplitude $\frac{a_2}{a_1}$ and relative frequency difference $\frac{\vert f_2 - f_1 \vert}{f_1}$.
We fixed $f_2 < f_1$ without loss of generality.
As expected, we observe that, for $a_2 \approx a_1$ and a relative frequency difference between $\frac{Q}{f_1 T}$ and $\frac{1}{Q}$, second-layer wavelets $\bm{\psi_{\lambda_2}}$ resonate with the difference tone as a result of the interference between signals $\bm{y_1}$ and $\bm{y_2}$.

\section{Application to manifold learning \label{sec:manifold}}
\begin{figure}
    \centering
    \includegraphics[width=0.85\linewidth]{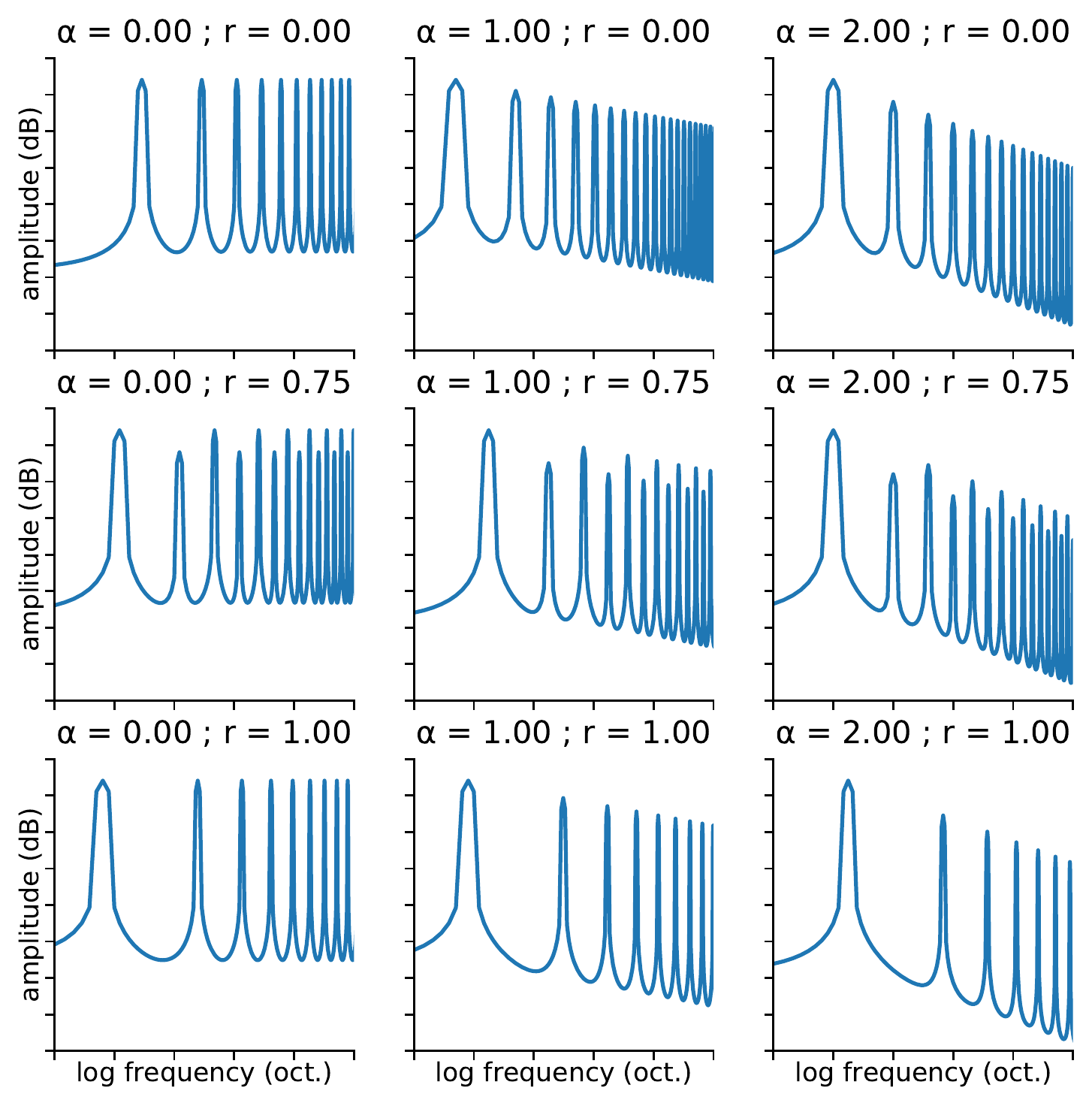}
    \caption{
    Log-magnitudes of synthetic musical tones as a function of wavelet log-frequency ($\log \lambda$).
    Ticks of the vertical (resp. horizontal) axis denote relative amplitude (resp. frequency) intervals of 10 dB (resp. one octave).
    Parameters $\alpha$ and $r$ denote the Fourier decay exponent and the relative odd-to-even amplitude difference $r$ respectively.
    See Equation 8 for details.}
    \label{eq:spectra}
\end{figure}

To demonstrate the ability of the scattering transform to characterize auditory masking, we build a dataset of complex tones according to the following additive synthesis model:
\begin{equation}
    \boldsymbol{y}_{\alpha,r}(t) =
    \sum_{n=1}^{N}
    \dfrac{
    1 + (-1)^{n} r
    }{
    n^{\alpha}
    }
    \cos(n f_1 t)
    \boldsymbol{\phi}_T(t),
    \label{eq:musical-tone}
\end{equation}
where $\boldsymbol{\phi}_T$ is a Hann window of duration $T$.
This additive synthesis model depends upon two parameters: the Fourier decay $\alpha$ and the relative odd-to-even amplitude difference $r$.
Figure \ref{eq:spectra} displays the CQT log-magnitude spectrum of $\boldsymbol{y}_{\alpha,r}$ for different values of $\alpha$ and $r$.
In practice, we set $T$ to $1024$ samples, $N$ to $32$ harmonics, and $f_1$ between $12$ and $24$ cycles.

Our synthetic dataset comprises $2500$ audio signals in total, corresponding to $50$ values of $\alpha$ between $0$ and $2$ and $50$ values of $r$ between $0$ and $1$, while $f_1$ is an integer chosen uniformly at random between $12$ and $24$.
We extract the scattering transform of each signal $\boldsymbol{y}_{\alpha,r}$ up to order $M=2$, with $Q=1$ and $J=8$, by means of the Kymatio Python package \cite{andreux2020jmlr}.
Concatenating $QJ$ first-order coefficients with $\frac{1}{2}Q^2 J(J-1)$  second-order coefficients yields a representation in dimension $37$.

For visualization purposes, we bring the $37$-dimensional space of scattering coefficients to the dimension three by means of the Isomap algorithm for unsupervised manifold learning \cite{tenenbaum2000science}.
The appeal behind Isomap is that pairwise Euclidean distances in the 3-D point cloud approximate the corresponding geodesic distances over the $K$-nearest neighbor graph associated to the dataset.
Throughout this paper, we set the number of neighbors to $K=100$ and measure neighboring relationships by comparing high-dimensional $\ell^2$ distances.
Crucially, in the case of the scattering transform, these $\ell^2$ distances are provably stable (i.e., Lipschitz-continuous) to the action of diffeomorphisms \cite[Theorem 2.12]{mallat2012cpam}.

\begin{figure}[t]
    \begin{center}
        \small{Scattering transform embedding}
        \vspace*{-0.4cm}
    \end{center}
    \begin{center}
    \hspace*{-0.8cm}
    \includegraphics[width=0.4\linewidth]{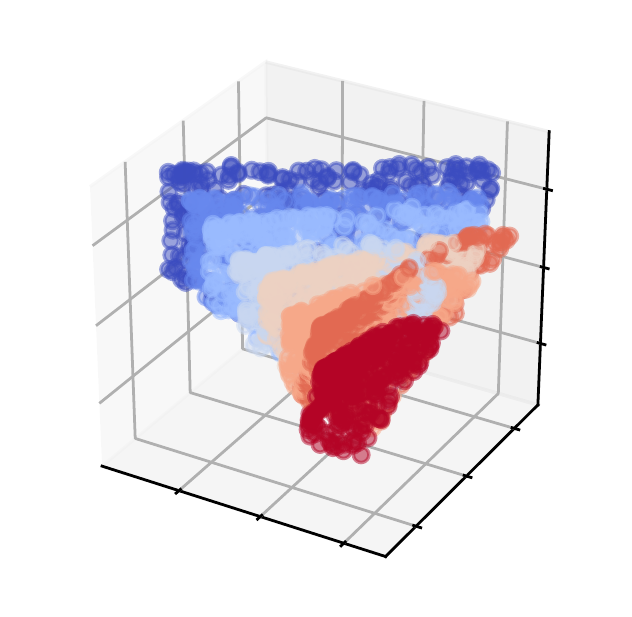}
    \hspace*{-0.8cm}
    \includegraphics[width=0.4\linewidth]{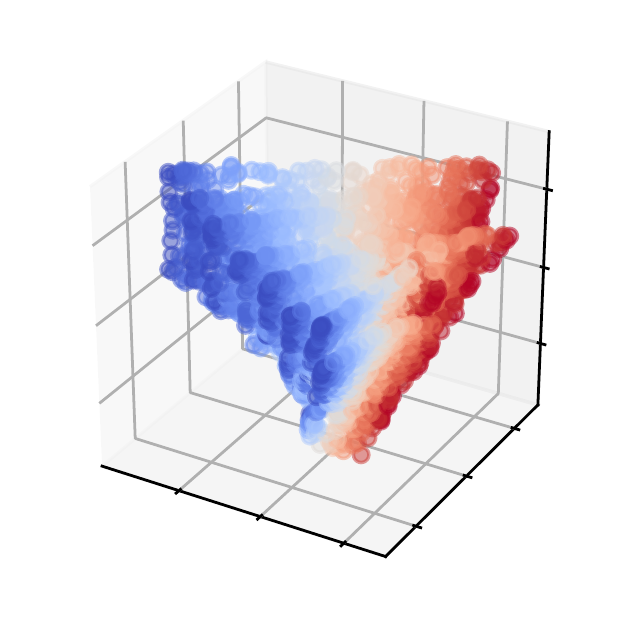}
    \hspace*{-0.8cm}
    \includegraphics[width=0.4\linewidth]{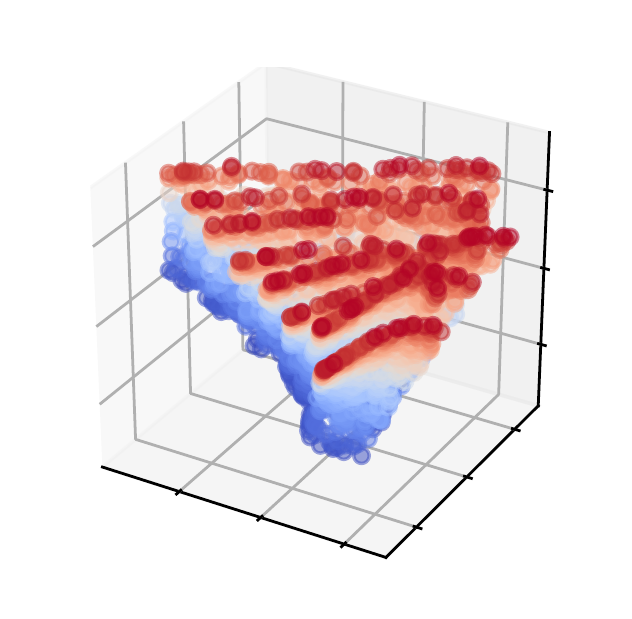}
    \end{center}
    \begin{center}
        \small{Audiovisual correspondence (Open-L3) embedding}
        \vspace*{-0.4cm}
    \end{center}
    \begin{center}
    \hspace*{-0.8cm}
    \includegraphics[width=0.4\linewidth]{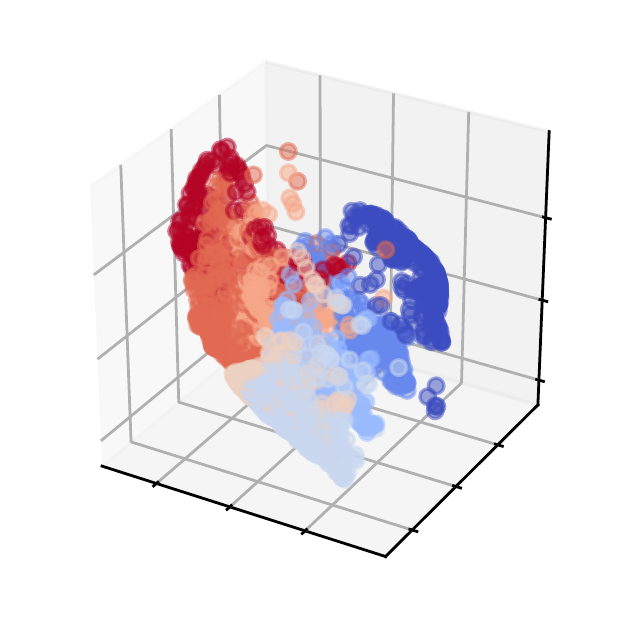}
    \hspace*{-0.8cm}
    \includegraphics[width=0.4\linewidth]{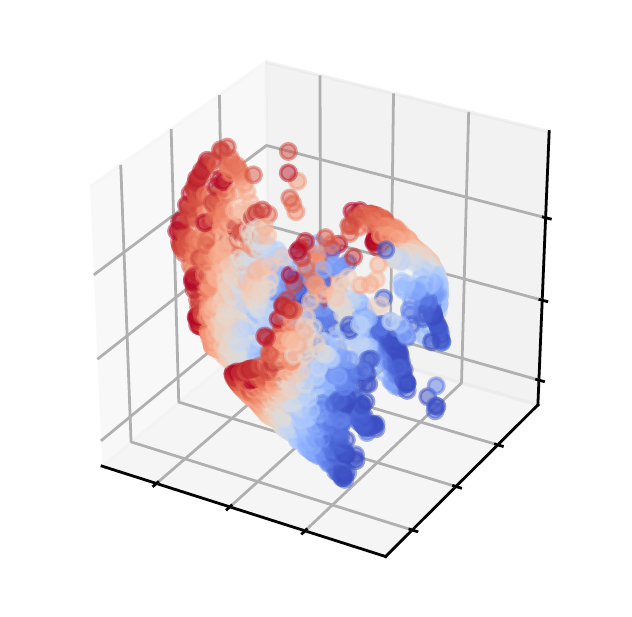}
    \hspace*{-0.8cm}
    \includegraphics[width=0.4\linewidth]{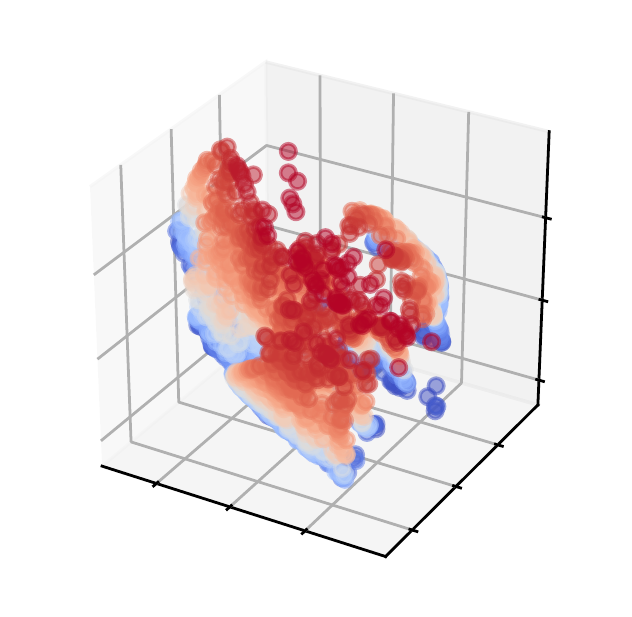}
    \end{center}
    \begin{center}
        \small{Mel-frequency cepstral coefficient (MFCC) embedding}
        \vspace*{-0.4cm}
    \end{center}
    \begin{center}
    \hspace*{-0.8cm}
    \includegraphics[width=0.4\linewidth]{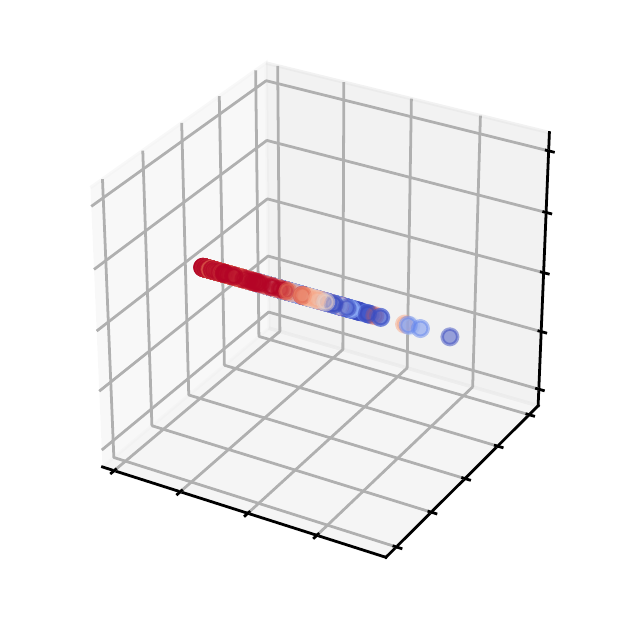}
    \hspace*{-0.8cm}
    \includegraphics[width=0.4\linewidth]{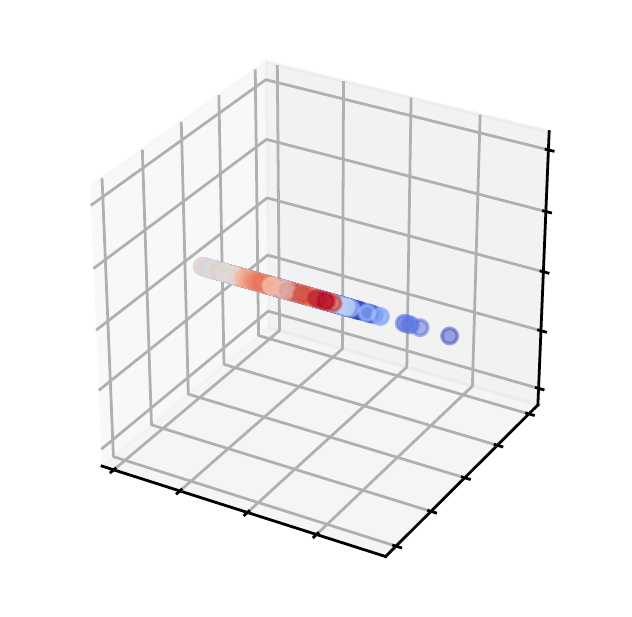}
    \hspace*{-0.8cm}
    \includegraphics[width=0.4\linewidth]{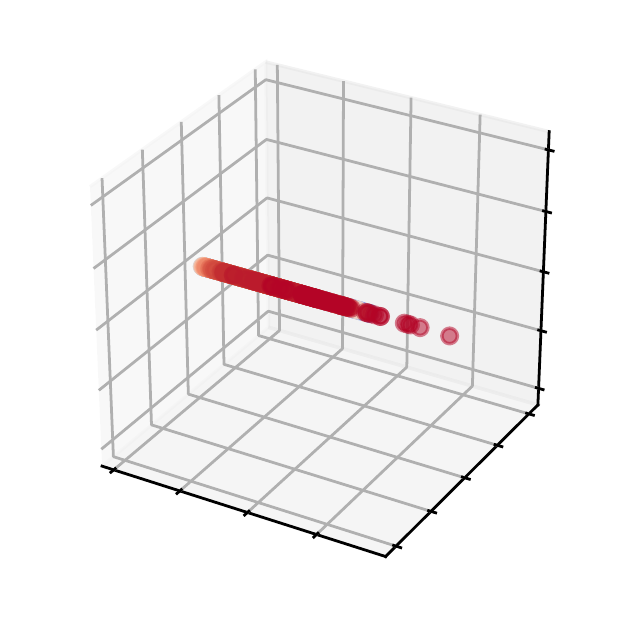}
    \end{center}
    \begin{center}
        \vspace*{-0.4cm}
        \small{Color: $f_1 \qquad\qquad$
        Color: $\alpha \quad\qquad\qquad$
        Color: $r\qquad$}
    \end{center}
    \caption{Isomap embedding of synthetic musical notes, as described by their scattering transform coefficients (top); their Open-L3 coefficients (center); and their mel-frequency cepstral coefficients (MFCC, bottom).
    The color of a dot, ranging from red to blue via white, denotes the fundamental frequency $f_1$ (left), the Fourier decay exponent $\alpha$ (center), and the relative odd-to-even amplitude difference $r$ (right) respectively.
    Note that all methods are unsupervised: triplets ($f_1$, $\alpha$, $r$) are not directly supplied to the models, but only serve for color grading post hoc.
    See Section \ref{sec:manifold} for details.}
    \label{fig:isomap}
\end{figure}

Figure \ref{fig:isomap} (top) illustrates our findings.
We observe that, after scattering transform and Isomap dimensionality reduction, the dataset appears as a 3-D Cartesian mesh whose principal components align with $f_1$, $\alpha$, and $r$ respectively.
This result demonstrates that the scattering transform is capable of disentangling and linearizing multiple factors of variability in the spectral envelope of periodic signals, even if those factors are not directly amenable to diffeomorphisms.

As a point of comparison, Figure \ref{fig:isomap} presents the outcome of Isomap on alternative feature representations: Open-L3 embedding (center) and mel-frequency cepstral coefficients (MFCC, bottom).
The former results from training a deep convolutional network (convnet) on a self-supervised task of audiovisual correspondence, and yields $6177$ coefficients \cite{cramer2019icassp}.
The latter resuts from a log-mel-spectrogram representation, followed by a discrete cosine transform (DCT) over the mel-frequency axis, and yields $12$ coefficients.
We compute MFCC with librosa v0.7 \cite{mcfee2020librosa} default parameters.

We observe that Open-L3 embeddings correctly disentangles boundary conditions ($r$) from fundamental frequency ($f_1$), but fails to disentangle Fourier decay ($\alpha$) from $f_1$.
Instead, correlations between $r$ and $f_1$ are positive for low-pitched sounds ($12$ to $16$ cycles) and negative for high-pitched sounds ($16$ to $24$  cycles).
Although this failure deserves a more formal inquiry, we hypothesize that this it stems from the small convolutional receptive field of the $L^{3}$-Net: $24$ mel subbands, i.e., roughly half an octave around $1$ kHz.

\begin{figure}
    \begin{center}
    \includegraphics[width=0.75\linewidth]{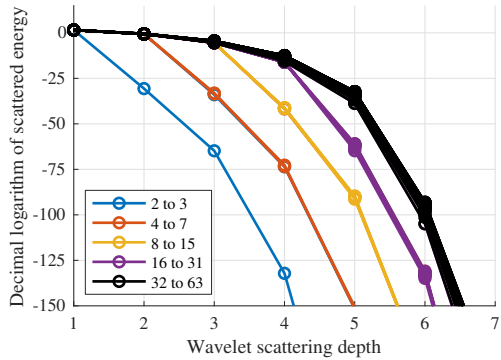}
    \end{center}
    \caption{Energy decay as a function of wavelet scattering depth $m$, for mixtures of $N$ components with equal amplitudes, equal phases, and evenly spaced frequencies.
    The color of each line plot denotes the integer part of $\log_2 N$.
    In this experiment, wavelets have a sine cardinal profile (Shannon wavelets) and a quality factor equal to $Q=1$. Each filterbank covers seven octaves.}
    \label{fig:decay}
\end{figure}

Moreover, in the case of MFCC, we find that the variability in fundamental frequency ($f_1$) dominates the variability in spectral shape parameters ($\alpha$ and $r$), thus yielding a rectilinear embedding (top).
This observation is in line with a previous publication \cite{lostanlen2016ismir}, which showed statistically that MFCCs are overly sensitive to frequency transposition in complex tones.

From this qualitative benchmark, it appears that the scattering transform is a more interpretable representation of periodic signals than Open-L3, while incurring a smaller computational cost.
However, in the presence of aperiodic signals such as environmental sounds, Open-L3 outperforms the scattering transform in terms of classification accuracy with linear support vector machines  \cite{arandjelovic2017look}.
To remain competitive, the scattering transform must not only capture heterodyne interference, but also joint spectrotemporal modulations \cite{anden2019joint}.
In this context, future work will strive to combine insights from multiresolution analysis and deep self-supervised learning.

\section{Beyond pairwise interference:\\ full-depth scattering networks}

In speech and music processing, pitched sounds are rarely approximable as a mixture of merely two components.
More often than not, they contain ten components or more, and span across multiple octaves in the Fourier domain.
Thus, computing the masking coefficient at the second layer only provides a crude description of the timbral content within each critical band.
Indeed, $\mathbf{S_2}$ encodes pairwise interference between sinusoidal components but fails to characterize more intricate structures in the spectral envelope of $\bm{y}$.

To address this issue, we propose to study the scattering transform beyond order two, thus encompassing heterodyne structures of greater multiplicity.
For the sake of mathematical tractability, we consider the following mother wavelet, hereafter called ``complex Shannon wavelet'' after \cite[Section 7.2.2]{mallat2008book}:
\begin{equation}
    \boldsymbol{\psi}:t \longmapsto
    \dfrac{
        \exp(2\mathrm{i}t) - \exp(\mathrm{i}t)
    }{
        2\pi\mathrm{i} t
    }
    \label{eq:complex-shannon}
\end{equation}
The definition of a scattering transform with complex Shannon wavelets requires to resort to the theory of tempered distributions.
We refer to \cite{strichartz2003book} for further mathematical details.

The following theorem, proven in the Appendix, describes the response of a deep scattering network in the important particular case of a periodic signal with finite bandwidth.
\begin{thm}
\label{thm:decay}
Let $\boldsymbol{y} \in \mathcal{C}^{\infty}(\mathbb{R})$ a periodic signal of fundamental frequency $f_1$.
Let $\boldsymbol{\psi}$ the complex Shannon wavelet as in Equation \ref{eq:complex-shannon} and $\mathbf{U_1}$ its associated scalogram operator as in Equation \ref{eq:scalogram}.
If $\boldsymbol{y}$ has a finite bandwidth of $M$ octaves, then its scattering coefficients $\mathbf{U_m}\boldsymbol{y}$ are zero for any $m>M$.
\end{thm}
This result is in agreement with the theorem of exponential decay of scattering coefficients \cite{waldspurger2017exponential}.
Note, however, that \cite{waldspurger2017exponential} expresses an upper bound on the energy at fixed depth for integrable signals, while we express an upper bound on the depth at fixed bandwidth for periodic signals.

We apply the theorem above to the case of a signal containing $N$ components of equal amplitudes, equal phases, and evenly spaced frequencies: $\boldsymbol{y}:t\mapsto\sum_{n=1}^{N} a_1 \cos(n f_1 t + \varphi_1)$.
Figure \ref{fig:decay} illustrates the decay of scatterered energy as a function of depth.
The conceptual analogy between depth and scale was originally proposed by \cite{mallat2016understanding} in a theoretical effort to clarify the role of hierarchical symmetries in convnets.

Although our findings support this analogy, we note that computing a scattering transform with $M=\log_2 T$ layers is often impractical.
However, if the Fourier series in $\boldsymbol{y}$ satisfies a self-similarity assumption, it is possible to match the representational capacity of a full-depth scattering network while keeping the depth to $M=2$.
Indeed, spiral scattering performs wavelet convolutions over time, over log-frequency, and across octaves, thereby capturing the spectrotemporal periodicity of Shepard tones and Shepard-Risset glissandos \cite{lostanlen2015dafx}.
Further research is needed to integrate broadband demodulation into deep convolutional architectures for machine listening.

\section{Conclusion}
In this article, we have studied the role of every layer in a scattering network by means of a well-established methodology, colloquially known as ``one or two components'' \cite{rilling2008tsp}.
We have come up with a numerical criterion of psychoacoustic masking; demonstrated that the scattering transform disentangles multiple factors of variability in the spectral envelope; and proven that the effective scattered depth of Fourier series is bounded by the logarithm of its bandwidth, thus emphasizing the importance of capturing geometric regularity across temporal scales.
\newpage

\printbibliography

\section*{Appendix: proof of Theorem \ref{thm:decay}}
\begin{proof}
We reason by induction over the depth variable $M$.
The base case ($M=1$) leads to $\mathbf{U_1}\boldsymbol{y}(t,\lambda) = 1$ if $\lambda < f_1 \leq 2\lambda$ and zero otherwise.
Because $\boldsymbol{\psi}$ has one vanishing moment, it follows that $\mathbf{U_2}\boldsymbol{y}$ is zero, and likewise at deeper layers.
To prove the induction step at depth $M$, to decompose $\boldsymbol{y}$ into a low-pass approximation $(\bm{y}\ast\bm{g_M})$ spanning the subband $[0; 2^{M} f_1[$ and a high-pass detail $(\bm{y}\ast\bm{h_M})$ spanning the subband $[2^{M} f_1; 2^{(M+1)} f_1[$.
Denoting by $c_n$ the complex-valued Fourier coefficients of $\boldsymbol{y}$, we have at every time $t\in\mathbb{R}$:
\begin{align}
    \boldsymbol{y}(t)
    &=
    (\boldsymbol{y}\ast\boldsymbol{g_M})(t) &+&
    (\boldsymbol{y}\ast\boldsymbol{h_M})(t)
    \nonumber \\
    &=
    \sum_{\vert n \vert \leq 2^M} c_n \exp(\mathrm{i} n f_1 t) &+&
    \sum_{\vert n \vert > 2^M} c_n \exp(\mathrm{i} n f_1 t)
\end{align}
On one hand, the coarse term $(\boldsymbol{y}\ast\boldsymbol{g_M})$ has a bandwidth of $M$ octaves.
Therefore, by the induction hypothesis, we have $\mathbf{U_m}(\boldsymbol{y}\ast\boldsymbol{g_M}) = 0$ for $m>M$, and \emph{a fortiori} for $m > (M+1)$.
On the other hand, we consider the complex Shannon scalogram of  $(\boldsymbol{y}\ast\boldsymbol{h_M})$ in some subband $\lambda>0$:
\begin{multline}
    \big\vert\boldsymbol{y}\ast\boldsymbol{h_M}\ast\boldsymbol{\psi}_{\lambda}\big\vert^2(t) \leq
    \big\vert\boldsymbol{y}\ast\boldsymbol{h_M}\ast\boldsymbol{\psi}_{2^M}\big\vert^2 (t) \hfill \\ \hfill =
    \sum_{n=(1+2^M)}^{2^{M+1}}
    \sum_{k=(1+2^M)}^{2^{M+1}}
    c_n c_k^{\ast} \exp\big(\mathrm{i} (n-k) f_1 t\big)
\end{multline}
In the double sum above, all integer differences of the form $(n-k)$ range between $-(2^{M}-1)$ and $(2^{M}-1)$.
Thus, $\big\vert\boldsymbol{y}\ast\boldsymbol{h_M}\ast\boldsymbol{\psi}_{2^M}\big\vert^2$ is a periodic signal of fundamental frequency $f_1$ spanning $M$ octaves.
Furthermore, because $\boldsymbol{h_M} = \boldsymbol{\psi}_{2^M}$,  $\big\vert\boldsymbol{y}\ast\boldsymbol{h_M}\ast\boldsymbol{\psi}_{\lambda}\big\vert^2$ has a smaller bandwidth than $\big\vert\boldsymbol{y}\ast\boldsymbol{h_M}\ast\boldsymbol{\psi}_{2^M}\big\vert^2$; i.e., $M$ octaves or less.
By the induction hypothesis, we have:
\begin{equation}
    \forall \lambda,\;
    \mathbf{U_{M+1}}\big(\vert\boldsymbol{y}\ast\boldsymbol{h_M}\ast\boldsymbol{\psi}_{\lambda}\vert^2\big)(\lambda_1,\ldots,\lambda_{m+1}) = 0.
\end{equation}
In the equation above, we recognize the scattering path $p=(\lambda,\lambda_1,\ldots,\lambda_{M+1})$ of $\mathbf{U_{M+2}}$.
Finally, because the scattering transform is a nonexpansive operator \cite[Prop. 2.5]{mallat2012cpam}, we have the inequality:
\begin{equation}
    \big\Vert \mathbf{U_{M+2}}(\boldsymbol{y}) \big\Vert
    \leq
    \big\Vert \mathbf{U_{M+2}}(\boldsymbol{y}\ast\boldsymbol{g_M}) \big\Vert
    +
    \big\Vert \mathbf{U_{M+2}}(\boldsymbol{y}\ast\boldsymbol{h_M}) \big\Vert = 0,
\end{equation}
which implies $\mathbf{U_{M+2}}\boldsymbol{y} = 0$, and likewise at deeper layers.
We conclude by induction that the theorem holds for any $M$.
\end{proof}

\end{document}